\documentclass[prb,showpacs,12pt]{revtex4}%
\usepackage{amsmath}
\usepackage{amssymb}
\usepackage{graphicx}
\usepackage{bm}
\usepackage{amsfonts}%
\newcommand{\be}{\begin{eqnarray}}
\newcommand{\en}{\end{eqnarray}}
\newcommand{\ben}{\begin{eqnarray}}
\newcommand{\enn}{\end{eqnarray}}
\newcommand{\beq}{\begin{eqnarray}}
\newcommand{\eeq}{\end{eqnarray}}

\begin{document}

\preprint{APS/123-QED}

\title{Magnetic hysteresis in a molecular Ising ferrimagnet:  Glauber  dynamics approach}

\author{A.~A.~Bukharov$^1$,
A.~S.~Ovchinnikov$^1$,
N.~V.~Baranov$^{1,2}$,
K.~Inoue$^{3}$
}

\address{
$^1$Department of Physics, Ural State University, Ekaterinburg, 620083 Russia\\
$^2$Institute of Metal Physics RAS and Ural State University, Ekaterinburg, 620083, Russia\\
$^3$Institute for Advanced Materials Research, Hiroshima University, Hiroshima, Japan
}

\date{\today}

\begin{abstract}

Motivated by recent experimental results reporting giant coercive fields in Co(II)-based molecular magnets we present a 
theory of hysteresis  phenomena based on the Glauber stochastic dynamics. Unusual form of hysteresis loops is similar to  those of found in Co-based quasi-one-dimensional ferrimagnet CoPhOMe at low temperatures. Temperature dependence of the coercive field has a characteristic form with an inflection that  may serve as an indicator of the Glauber dynamics in real compounds.  A relevance of the model for other Co-based molecular magnets is discussed.
\end{abstract}

\pacs{Valid PACS appear here}
\maketitle

\section{Introduction}

One of the remarkable phenomena found in  molecular magnetic materials  is the magnetic hysteresis similar to those observed  in hard magnets.\cite{Novak,Bogani} In particularly, magnetic hysteresis can be observed in the absence of a long-range magnetic order in zero-dimensional  single-molecule magnets (SMM)\cite{Sessoli1} and one-dimensional (1D) single chain magnets (SCM).\cite{Caneschi,Clerac} The hysteresis phenomena in SMM are affected by  large easy-axis magnetic
anisotropy and by the weak intermolecular interactions.\cite{Villain} Although  a large easy-axis
magnetic anisotropy can also be of importance in quasi-1D magnets,  an origin of magnetic hysteresis in these systems  is not yet well understood.\cite{Sessoli}  

Discovery of slow relaxation of magnetization in quasi-1D compound Co(hfac)$_2$[NIT(C$_6$H$_4p$-OMe)] (or CoPhOMe)\cite{Caneschi} evoked interest to the Glauber kinetic  model suggested  for relaxation dynamics in 1D Ising ferromagnet.\cite{Glauber,Kawasaki,Suzuki} The source of a strong easy-axis anisotropy are Co(II) ions that enable to treat collective properties of this compound by means of 1D  Ising model. Recently, pronounced hysteresis loops with giant coercivity were observed in another Co(II)-based materials, namely [Co(hfac)$_2$NIT-C$_6$H$_4$-O-R]\cite{Ishii}, where  R=(CH$_2$)$_3$CH$_3$, and [Co(hfac)$_2$]$\cdot$BNO$^\ast$ (BNO$^\ast$ is chiral triplet bis(nitroxide) radical).\cite{Baranov}  Despite the Glauber dynamics seems to be plausible, one should take into account that  chains are packed in a 3D structure, and with a cooling 1D units undergo a phase
transition into 3D magnetic array. It means that another explanation of hysteresis  phenomena related with domain wall (DW) dynamics comes into the play. High coercive fields reported in Refs.\cite{Ishii,Baranov}  at low temperatures, i.e. broad hysteresis loops, support the reasonings. Note that considerations of slow magnetic relaxation below 3D ordering in the framework of the DW approach turns out to be fruitful  for  Mn(II)-based quasi-1D ferrimagnets.\cite{Ovchinnikov,Lhotel}

Being guided this motivation we present a theory of magnetic hysteresis for Ising ferrimagnetic chain calculated within the Glauber dynamics approach. An extension of  the Glauber model for higher spins (more than $1/2$)  has been done in a modeling  of photoinduced reversible magnetization \cite{Nishino2001,Nishino2005}. Despite equilibrium properties of mixed spin Ising systems have attracted a much of attention, a study of  nonequilibrium aspects of the model started only recently.\cite{Deviren}   In the paper we perform the mean field treatment of the Glauber-type stochastic dynamics  of the mixed (3/2,1) ferrimagnetic chain by considering  fluctuations of local fields in the spirit of the generalized mean-field theory.\cite{Meilikhov} Our goal is to verify a validity of the model for real Co-based molecular ferrimagnetic chains. The choice of the spin  quantum numbers is stipulated by a possible prototype of the model, the metal-organic compound  [Co(hfac)$_2$]$\cdot$BNO$^{\ast}$ studied recently by two of us, which embodies  Co(II) ions with spin 3/2 and the chiral triplet biradical ligands  BNO$^\ast$   with spin 1. However, the results we obtain may be of interest for  another Co-based quasi-1D magnets. 

The paper is organized as follows. In Sec.II we  obtain  dynamic equations of the Ising ferrimagnetic ($S$,$\sigma$) chain  model  derived through the Glauber stochastic dynamics in a presence of a time-varying magnetic field. In Sec. III we present  results of  hysteresis behavior obtained  via  numerical calculations of these equations for the case $S=3/2$, $\sigma=1$. A dependence of  shape  of the hysteresis cycles on the ratio between the magnetic field frequency and the spin flip frequency is studied. In Sec. IV we confirm by a Monte Carlo (MC) method that a hysteresis loops arise  in a static magnetization process.  Discussions are relegated to the Conclusion part.

\section{Model}

The Hamiltonian of the Ising (3/2,1) ferrimagnetic chain is given by
\begin{equation}
{\cal H} = J \sum_i \left(  \sigma_i S_i + \sigma_{i+1} S_i \right) - H(t) \left(  \sum_i \sigma_i + \sum_ i S_i \right), 
\label{Hamiltonian}
\end{equation}
where the first term  sums interactions between  the nearest neighbors with the spin variables $\sigma_i =0,\pm1$ and $S_i = \pm 1/2, \pm 3/2$, and $J > 0$ favors an antiferromagnetic coupling  of the adjacent sites. The Zeeman term describes interaction of the spins with an oscillating magnetic field of the sinusoidal form
$$
H(t)=H_0 \cos(\omega t)
$$
with a frequency $\omega$. The system is in contact with an isothermal heat bath at given temperature $T$.

The relaxation of the interacting Ising system cannot be obtained from its Hamiltonian because it eliminates intrinsic spin dynamics (precession in the local field). Nevertheless, relaxation phenomena can be described by means of a phenomenological equation which specifies the transition rate from one spin configuration to another. In general, we assume that transition from one configuration to another involves changing a single spin, i.e. according to Glauber dynamics,\cite{Glauber,Kawasaki,Suzuki} at a rate of $1/\tau$ transitions per unit time.

According to formalism of Ref.\cite{Deviren} the master equation for one sublattice can be written in suggestion that  spins on another sublattice momentarily frozen, i.e.
\begin{multline}
\frac{d}{dt} P_{\{ S\}}(\sigma_1,\sigma_2,\ldots, \sigma_j, \ldots, \sigma_N) 
\\{} = 
 \sum_j \sum_{\sigma'_j \neq \sigma_j} \omega_j(\sigma'_j \rightarrow \sigma_j) P_{\{ S\}}(\sigma_1,\sigma_2,\ldots, \sigma'_j, \ldots, \sigma_N)
\\{} - 
 \sum_j \left( \sum_{\sigma'_j \neq \sigma_j} \omega_j(\sigma_j \rightarrow \sigma'_j) \right) P_{\{ S\}}(\sigma_1,\sigma_2,\ldots, \sigma_j, \ldots, \sigma_N) 
 , 
\label{sigma's ME}
\end{multline}
\begin{multline}
\frac{d}{dt} P_{\{ \sigma\}}(S_1,S_2,\ldots, S_j, \ldots, S_N) 
\\{} =
 \sum_j \sum_{S'_j \neq S_j} \omega_j(S'_j \rightarrow S_j) P_{\{ \sigma\}}(S_1,S_2,\ldots, S'_j, \ldots, S_N)
\\{} -
 \sum_j \left( \sum_{S'_j \neq S_j} \omega_j(S_j \rightarrow S'_j) \right) P_{\{ \sigma\}}(S_1,S_2,\ldots, S_j, \ldots, S_N) 
 , 
\label{S's ME}
\end{multline}
where $P_{\{ S\}}(\sigma_1,\sigma_2,\ldots \sigma_j, \ldots, \sigma_N)$  is the probability that the system has the spin configuration $\{\sigma_1,$ $\sigma_2,$ $\ldots$ $\sigma_j,$ $\ldots,$ $\sigma_N \}$ in the first sublattice leaving  the spins $\{ S \}$ of the second sublattice fixed, $\omega_j(\sigma'_j \rightarrow \sigma_j)$ is the probability per unit time that the $j$-th spin changes from the value $\sigma'_j$ to $\sigma_j$. The similar notations hold for another sublattice.

The transition probabilities $\omega_j(\sigma'_j \rightarrow \sigma_j)$ and $\omega_j(S'_j \rightarrow S_j)$ are imposed by the principle of detailed balance. Indeed, the principle requires the probabilities of states $P_{\{ S\}}$ and $P_{\{ \sigma\}}$  to be stationary at equilibrium
$$
\omega_j(\sigma_j \rightarrow \sigma'_j) P_{\{ S\}}(\sigma_1,\sigma_2,\ldots, \sigma_j, \ldots, \sigma_N)
 =
\omega_j(\sigma'_j \rightarrow \sigma_j) P_{\{ S\}}(\sigma_1,\sigma_2,\ldots, \sigma'_j, \ldots, \sigma_N),
$$
$$
\omega_j(S_j \rightarrow S'_j) P_{\{ \sigma\}}(S_1,S_2,\ldots, S_j, \ldots, S_N)
 =
\omega_j(S'_j \rightarrow S_j) P_{\{ \sigma\}}(S_1,S_2,\ldots, S'_j, \ldots, S_N).
$$
Provided the probabilities $P_{\{ S\}}$ and $P_{\{ \sigma\}}$  are Boltzmann distributions, i.e.
\begin{equation}
\frac{
P_{\{ S\}}(\sigma_1,\sigma_2,\ldots, \sigma'_j, \ldots, \sigma_N) 
}{ 
P_{\{ S\}}(\sigma_1,\sigma_2,\ldots, \sigma_j, \ldots, \sigma_N) 
}=
\frac{
\exp\Bigl[ -\beta \Bigl\{ J \Bigl( \sigma'_j S_{j-1} + \sigma'_j S_j \Bigr) - 
H \sigma'_j \Bigr\} \Bigr] 
}{
\exp\Bigl[ -\beta \Bigl\{ J \Bigl( \sigma_j S_{j-1} + \sigma_j S_j \Bigr) - 
H \sigma_j \Bigr\} \Bigr] 
},
\end{equation}
\begin{equation}
\frac{
P_{\{ \sigma\}}(S_1,S_2,\ldots, S'_j, \ldots, S_N) 
}{ 
P_{\{ \sigma\}}(S_1,S_2,\ldots, S_j, \ldots, S_N) 
}
=
\frac{
\exp\Bigl[ -\beta \Bigl\{ J \Bigl( \sigma_j S'_j + \sigma_{j+1} S'_j \Bigr) - 
H S'_j \Bigr\} \Bigr] 
}{
\exp\Bigl[ -\beta \Bigl\{ J \Bigl( \sigma_j S_j + \sigma_{j+1} S_j \Bigr) - 
H S_j \Bigr\} \Bigr] 
}
\end{equation}
one get 
\begin{equation}
\frac{ \omega_j(\sigma_j \rightarrow \sigma'_j) }{ \omega_j(\sigma'_j \rightarrow \sigma_j) } 
= \frac{ \exp[ \sigma'_j y_j ] }{ \exp[ \sigma_j y_j ] }, 
\qquad
\frac{ \omega_j(S_j \rightarrow S'_j) }{ \omega_j(S'_j \rightarrow S_j) } 
= \frac{ \exp[ \xi_j S'_j ] }{ \exp[ \xi_j S_j ] },
\label{omega_relation}
\end{equation}
where $y_j = \beta \bigl[ H - J ( S_{j-1} + S_j ) \bigr]$, and $\xi_j = \beta \bigl[ H - J ( \sigma_j + \sigma_{j+1} ) \bigr]$, $\beta=1/T$.

In Glauber dynamics the relationships (\ref{omega_relation}) are resolved as follows
\begin{equation}
\omega_j(\sigma_j \rightarrow \sigma'_j)
= 
\Omega \frac{ \exp[ \sigma'_j y_j ] }{ \sum\limits_{\sigma''_j} \exp[ \sigma''_j y_j ] },
\qquad
\omega_j(S_j \rightarrow S'_j)
= 
\Omega \frac{ \exp[ \xi_j S'_j ] }{ \sum\limits_{S''_j} \exp[ \xi_j S''_j ] } 
,
\label{gl_freq}
\end{equation}
where $\Omega=1/\tau$ is a number of spin changes per unit time. Strictly speaking, Eqs.(\ref{gl_freq}) are relevant for a equilibrium process with a constant magnetic field. Nevertheless, it is reliable when the field sweep frequency   is much less then that  of spin transitions $\omega \ll \Omega$.

Expectation value of the $\sigma$-sublattice  magnetization at the moment $t$ is given by
 \begin{equation}
 \langle \sigma_i \rangle
= \sum_{\sigma_i} \sigma_i p(\sigma_i),
\label{expect}
\end{equation}
where  
\be
p(\sigma_i)
= 
\sum_{\{S\}} \sum_{\{ \sigma \}}{}^{'}  P_{\{S\}}(\sigma_1, \ldots, \sigma_i, \ldots, \sigma_N)
\label{prob}
\en
is the probability to find the $\sigma_i$ value. The sum with the prime runs over all $\sigma$ variables except that of $i$-th site. Similar definitions are hold for another sublattice.

The probability satisfies the following dynamic equation
\be
\frac{d p(\sigma_i)}{dt}
= -\Omega p(\sigma_i) + \Omega \Bigl\langle \frac{ \exp[\sigma_i y_i] }{ \sum\limits_{\sigma'_i} \exp[ \sigma'_i y_i] } \Bigr\rangle, 
\label{sigma_dt}
\en
which is obtained from Eq.(\ref{prob}) with an account of Eqs.(\ref{sigma's ME}-\ref{S's ME}) and (\ref{gl_freq}). The average in the right-hand side is determined as follows
\begin{equation}
\Bigl\langle \frac{ \exp[\sigma_i y_i] }{ \sum\limits_{\sigma'_i} \exp[ \sigma'_i y_i] } \Bigr\rangle 
= 
\sum_{\{S\}} \sum_{\{\sigma\}}  \frac{ \exp[\sigma_i y_i] }{ \sum\limits_{\sigma'_i} \exp[ \sigma'_i y_i] }  P_{\{ S\}}(\sigma_1, \ldots, \sigma_i, \ldots, \sigma_N).
\label{average}
\end{equation}
The analogous equation is derived for another sublattice
\begin{equation}
\frac{d p(S_i)}{dt}
= 
-\Omega  p(S_i) + \Omega \Bigl\langle \frac{ \exp[\xi_i S_i] }{ \sum\limits_{S'_i} \exp[ \xi_i S'_i] } \Bigr\rangle.
\label{general d/dt p}
\end{equation}
The system of Eqs.(\ref{sigma_dt},\ref{general d/dt p}) can be treated in the mean-field approximation (MFA)
\begin{equation}
\Bigl\langle \frac{ \exp[\sigma_i y_i] }{ \sum\limits_{\sigma'_i} \exp[ \sigma'_i y_i] } \Bigr\rangle 
\approx 
\frac{ \exp[\sigma_i \langle y_i \rangle ] }{ \sum\limits_{\sigma'_i} \exp[ \sigma'_i \langle y_i \rangle] },
\label{mf approx 1}
\end{equation}
that yields 
\begin{equation}
\left\{
\begin{array}{c}
\frac{d \langle \sigma_i \rangle}{dt}
= - \Omega \Bigl[
\langle \sigma_i \rangle
-\sigma B_{\sigma}(\sigma \langle y_i \rangle)
\Bigr], \\
{}\\
\frac{d \langle S_i \rangle}{dt}
= - \Omega \Bigl[
\langle S_i \rangle
- S B_{S}(S \langle \xi_i \rangle)
\Bigr],
\end{array}
\right.
\label{mf d/dt S}
\end{equation}
where the Brillouine function $B_s(x)$ is introduced
\begin{equation*}
B_s(x) = \Bigl( 1 + \frac1{2 s} \Bigr) \coth \Bigl[ \Bigl( 1 + \frac1{2 s} \Bigr) x \Bigr] 
- \frac1{2 s} \coth \Bigl[  \frac{x}{2 s} \Bigr].
\end{equation*}
By taking the uniform arrangement 
$\langle \sigma_i \rangle = \langle \sigma \rangle$,  $\langle S_i \rangle = \langle S \rangle$ and, as a consequence, 
$\langle y_j \rangle =  \beta [ H(t) - 2 J \langle S \rangle ]$, $\langle \xi_j \rangle 
= \beta [ H(t) - 2 J \langle \sigma \rangle ]$ one obtain eventually the dynamical equations
\begin{equation}
\left\{
\begin{array}{c}
\frac{d \langle \sigma \rangle}{dt}
= - \Omega \Bigl[
\langle \sigma \rangle -
\sigma B_{\sigma}\bigl(\sigma \beta \{ H(t) - 2 J \langle S \rangle \} \bigr)
\Bigr], \\
{}\\
\frac{d \langle S \rangle}{dt}
= - \Omega \Bigl[
\langle S \rangle -
S B_{S}\bigl(S \beta \{ H(t) - 2 J \langle \sigma \rangle \} \bigr)
\Bigr]. 
\end{array}
\right.
\label{homogen mf d/dt S}
\end{equation}
At equilibrium ($d/dt \to 0$) one recover the usual MFA equations from the system. 

\begin{figure}
	\centering
		\includegraphics{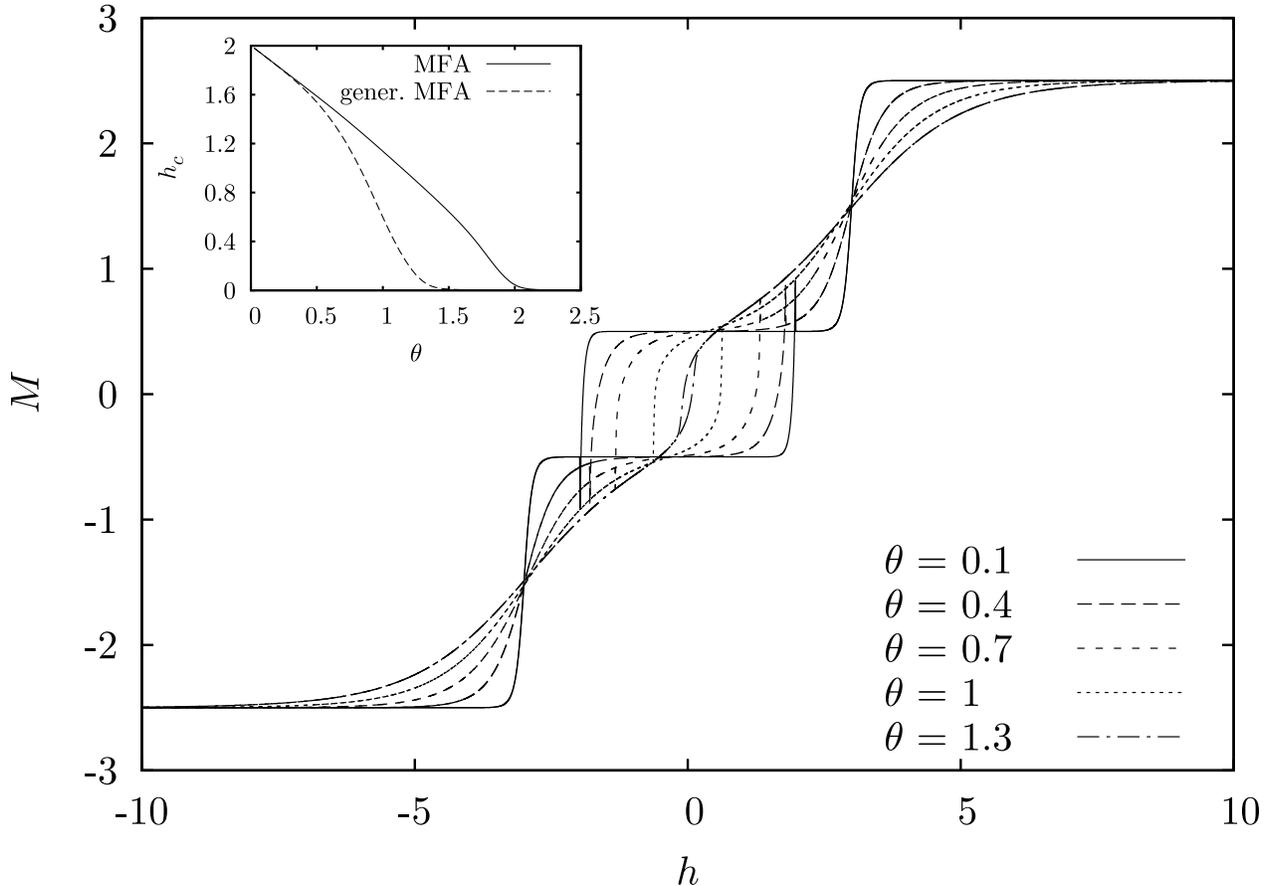}
	\caption{Hysteresis loops of magnetization per cell $M = \langle \sigma \rangle + \langle S \rangle$  at different  temperatures $\theta$. Inset: the coercive field $h_c$ vs temperature.}
	\label{fig:magnitizations}
\end{figure}

In one dimension the mean-field approximation is poor since the local fields $y_i$ and $\xi_i$ fluctuate strongly from one site to another. To overcome partly the drawback we use  a generalization of the MFA approach like those used in Ref.\cite{Meilikhov} 

Take the following approximation for the averages (\ref{average})
\begin{equation}
\Bigl\langle \frac{ \exp[\sigma_i y_i] }{ \sum\limits_{\sigma'_i} \exp[ \sigma'_i y_i] } \Bigr\rangle 
\approx 
\sum_{S_{i-1}, S_i} \frac{ \exp[\sigma_i y_i] }{ \sum\limits_{\sigma'_i} \exp[ \sigma'_i y_i] } p(S_{i-1}) p(S_i)
,
\label{mfg approx 1}
\end{equation}
\begin{equation}
\Bigl\langle \frac{ \exp[\xi_i S_i] }{ \sum\limits_{S'_i} \exp[ \xi_i S'_i] } \Bigr\rangle 
\approx 
\sum_{\sigma_i, \sigma_{i+1}} \frac{ \exp[\xi_i S_i] }{ \sum\limits_{S'_i} \exp[ \xi_i S'_i ] }  p(\sigma_i) p(\sigma_{i+1})
,
\label{mfg approx 2}
\end{equation}
i.e. the averages depend on the spin variables at the adjacent sites of another sublattice.

After substitution  Eqs.(\ref{mfg approx 1}-\ref{mfg approx 2}) into Eqs.(\ref{sigma_dt},\ref{general d/dt p}) one obtain the dynamic equations of the {\it generalized} mean-field approximation written through the probabilities
$$
\frac{d}{dt} p(\sigma_i)
= 
-\Omega \biggl[ p(\sigma_i) - \sum_{S_{i-1}, S_i} \frac{ \exp[\sigma_i y_i] }{ \sum\limits_{\sigma'_i} \exp[ \sigma'_i y_i] }\biggr|_{y_i = \beta [H - J ( S_{i-1} + S_i )]}  p(S_{i-1}) p(S_i) \biggr],
$$
\begin{equation}
\frac{d }{dt} p(S_i)
= 
-\Omega \biggl[ p(S_i) - \sum_{\sigma_i, \sigma_{i+1}} \frac{ \exp[\xi_i S_i] }{ \sum\limits_{S'_i} \exp[ \xi_i S'_i ] }\biggr|_{\xi_i = \beta [H - J ( \sigma_j + \sigma_{j+1} )]}  p(\sigma_i) p(\sigma_{i+1}) \biggr].
\label{mfg d/dt p}
\end{equation}
By determining a time evolution of these quantities one recover a time dependence of expectation values for the sublattice magnetizations according to Eq.(\ref{expect}).

\section{Numerical results}
Let us now discuss hysteresis phenomenon for $(3/2,1)$ ferrimagnetic Ising chain obtained on the base of Eqs.(\ref{homogen mf d/dt S}) and (\ref{mfg d/dt p}). For numerical calculations it is convenient to rewrite  the MFA equations in the form
$$
\frac{d \langle \sigma \rangle}{d {\tilde \tau}}
= - \langle \sigma \rangle
+ B_1 \left( [h - 2 \langle S \rangle ]/\theta \right),
$$
$$
\frac{d \langle S \rangle}{d {\tilde \tau}}
= - \langle S \rangle
+ (3/2) B_{\frac32}\left( 3[ h - 2 \langle \sigma \rangle ]/2 \theta \right).
\label{reduced homogen mf d/dt S}
$$
where ${\tilde \tau} = \Omega t$, and $h = H/J$, $\theta = T/J$ are reduced field and temperature, respectively.

Regarding the system (\ref{mfg d/dt p}) there is a way to bring down a number of differential equations from 7 till 5 by using the normalizing condition for probabilities $p$ and definition of observables. Indeed, hold four differential equations for the probabilities  $p(\pm1)$,  $p(\pm 3/2)$
\be
\frac{d}{d{\tilde \tau}} p(\pm 1)
= - p(\pm 1) + \sum_{S_{i-1}, S_i} \frac{ \exp[\pm ( h - S_{i-1} - S_i )/\theta ] }{ \sum\limits_{\sigma'_i} \exp[ \sigma'_i ( h - S_{i-1} - S_i )/\theta] }  p(S_{i-1}) p(S_i),
\label{redDE1}
\en
\be
\frac{d }{d{\tilde \tau}} p\left(\pm 3/2 \right)
= 
- p\left( \pm  3/2 \right) + \sum_{\sigma_i, \sigma_{i+1}} \frac{ \exp[\pm 3( h - \sigma_j - \sigma_{j+1} )/(2 \theta )] }{ \sum\limits_{S'_i} \exp[ S'_i \frac{ h - \sigma_j - \sigma_{j+1} }{ \theta } ] }  p(\sigma_i) p(\sigma_{i+1}),
\label{redDE2}
\en
and add the corresponding equation for the observable $\langle S \rangle$ 
$$
\frac{d \langle S \rangle}{d{\tilde \tau}}
= - \langle S \rangle + \sum_{\sigma_i, \sigma_{i+1}} \frac32 B_{\frac32}\Bigl(  3[ h - \sigma_j - \sigma_{j+1} ]/2  \theta \Bigr)  p(\sigma_i) p(\sigma_{i+1}),
$$
obtained from Eq.(\ref{mfg d/dt p}). The system is supplemented by the pure algebraic relations
$$
p(0) = 1 - p(-1) - p(1),
$$
$$
p(-1/2) = p(3/2) - 2 p(-3/2) - \langle S \rangle + 1/2,
$$
$$
p(1/2) = \langle S \rangle - 2 p(3/2) + p(-3/2) + 1/2
$$
that yield the leaving probabilities.

The sums in the right-hand sides of Eqs.(\ref{redDE1},\ref{redDE2}) are explicitly given by
$$
 \sum_{S_{i-1}, S_i} \frac{ \exp[\sigma_i ( h - S_{i-1} - S_i )/ \theta ] }{ \sum\limits_{\sigma'_i} \exp[ \sigma'_i ( h - S_{i-1} - S_i )/ \theta ] }  p(S_{i-1}) p(S_i)
= 
\frac{ \exp[\sigma_i ( h + 3 )/ \theta ] }{ Z_1[ ( h + 3 )/\theta ] }  p(-3/2)^2
$$
$$
+ 
2 \frac{ \exp[\sigma_i ( h + 2)/\theta ] }{ Z_1[ ( h + 2 )/ \theta ] }  p(-3/2) p(-1/2) 
+ 
\frac{ \exp[\sigma_i ( h + 1)/ \theta ] }{ Z_1[ ( h + 1 )/ \theta ] } \bigl[ p(-1/2)^2 + 2 p(-3/2) p(1/2) \bigr]
$$
$$
+ 
2 \frac{ \exp[\sigma_i  h/ \theta ] }{ Z_1[  h /\theta ] } \bigl[ p(-3/2) p(3/2) + p(-1/2) p(1/2) \bigr]
+ 
\frac{ \exp[\sigma_i ( h - 1)/\theta ] }{ Z_1[ ( h - 1)/\theta ] } \bigl[ p(1/2)^2 + 2 p(-1/2) p(3/2) \bigr]
$$
$$ 
+ 
2 \frac{ \exp[\sigma_i ( h - 2)/ \theta] }{ Z_1[ ( h - 2 )/\theta ] }  p(1/2) p(3/2) 
+ 
\frac{ \exp[\sigma_i ( h - 3)/ \theta ] }{ Z_1[ ( h - 3 )/ \theta ] }  p(3/2)^2,
$$
for the spin-1 sublattice and 
$$
\sum_{\sigma_i, \sigma_{i+1}} \frac{ \exp[S_i ( h - \sigma_j - \sigma_{j+1} )/\theta ] }{ \sum\limits_{S'_i} \exp[ S'_i ( h - \sigma_j - \sigma_{j+1} )/\theta ] }  p(\sigma_i) p(\sigma_{i+1})
= 
\frac{ \exp[S_i ( h + 2 )/ \theta ] }{Z_{\frac32}[( h + 2)/\theta] }  p(-1)^2 
$$
$$
+2 \frac{ \exp[S_i ( h + 1)/\theta ] }{Z_{\frac32}[( h + 1)/\theta]} p(-1) p(0) 
+ 
\frac{ \exp[S_i  h/\theta ] }{Z_{\frac32}[h/\theta]  } [ p(0)^2 + 2 p(-1) p(1) ] 
$$
$$ 
+2 \frac{ \exp[S_i ( h - 1 )/\theta ] }{ Z_{\frac32}[( h - 1)/\theta] } p(0) p(1) 
+ 
\frac{ \exp[S_i ( h - 2 )/ \theta ] }{ Z_{\frac32}[( h - 2)/\theta]}  p(1)^2,
$$
for the spin-3/2 one. Here, 
$$
Z_1( x ) = \sum\limits_{\sigma'_i} \exp[ \sigma'_i x] = 2 \cosh( x ) + 1, \qquad 
Z_{\frac32}( x ) = \sum\limits_{S'_i} \exp[ S'_i x ] = 2 \cosh(x/2) + 2 \cosh(3 x/2).
$$

The results of such calculations performed within the Runge-Kutta method are presented in Figs.1-3 for the field frequency $\omega/\Omega=10^{-4}$ ($\Omega=1$). Initial values were chosen to correspond either to total saturation  or disorder in both sublattices. A stationary regime is approximately reached  after one full sweep.   Fig.\ref{fig:magnitizations} shows the evolution of hysteresis loops at various temperatures. It can be clearly seen that the area of the hysteresis loop monotonically decreases with increasing temperature. The hysteresis curves are characterized by well-defined steps, and for higher temperature ($\theta>0.7$) have the form with a loop in the middle close to that of observed experimentally in Co(II)-based quasi-1D ferrimagnetic compound CoPhOMe at temperatures around $\gtrsim 4.5$ K.\cite{Caneschi} The process of thermally activated spin flip formation reduces the effective intrinsic coercive field (see inset in Fig.\ref{fig:magnitizations}). Note that the temperature dependence of the coercivity exhibits a behavior (a curve with an inflection) distinguished from that of predicted by a thermal activation theory by Egami\cite{Egami} used for hard magnetic materials.\cite{Parker} When $T$ tends to zero  the value of the coercive field becomes equal to $2\,J$, i.e. it is determined by the exchange coupling (not by anisotropy) as expected for Ising systems. 

The both analytical approaches give qualitatively same results (see Fig.2). Note only that a more detailed account of fluctuations within the generalized MFA squeezes the area of the hysteresis loops. 

\begin{figure}[h]
\begin{center}
\includegraphics[width=145mm]{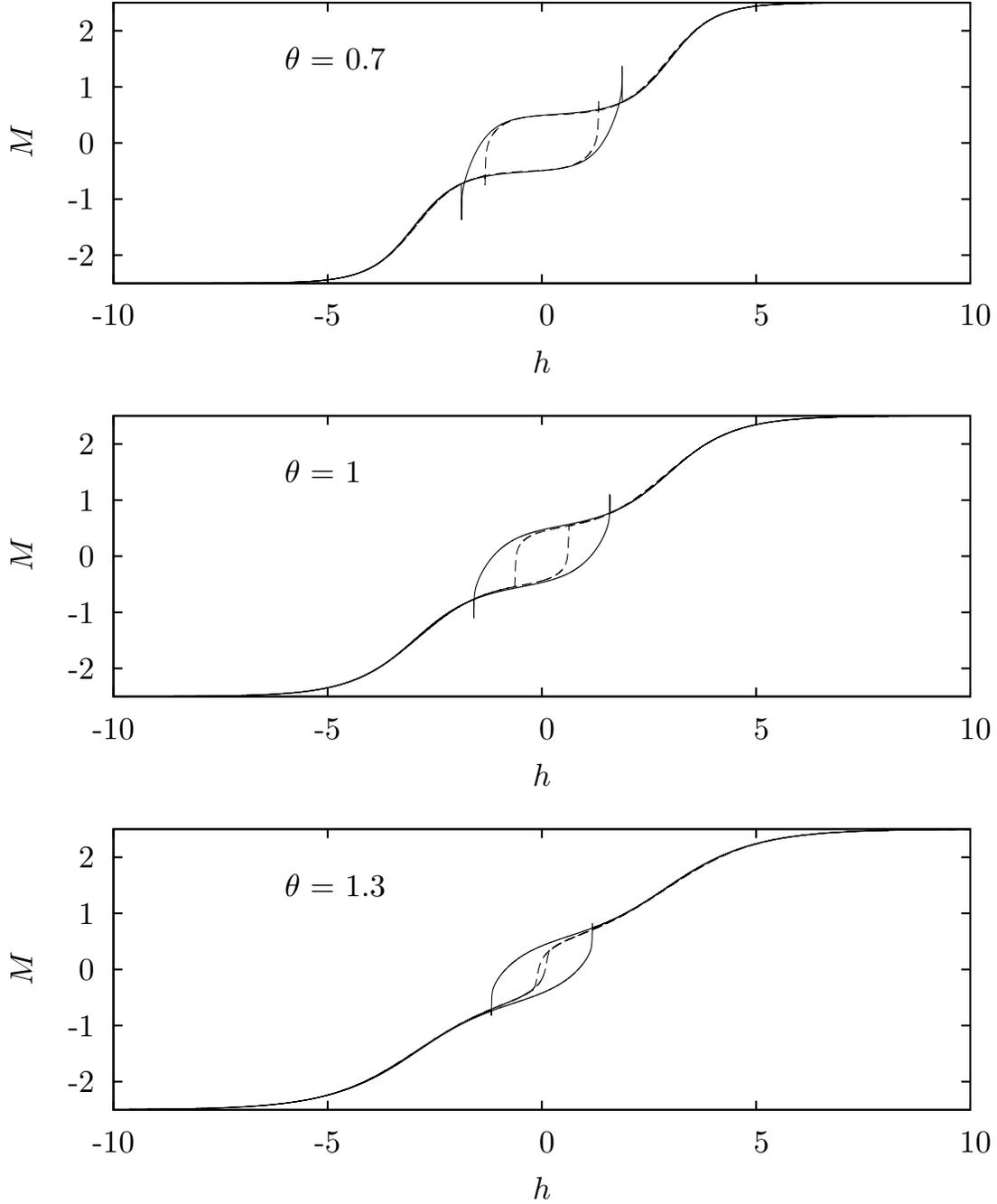}
\end{center}
	\caption{Comparison of hysteresis curves calculated within MFA (solid) and generalized MFA (dashed) approaches.}
	\label{fig:mf_vs_mfg}
\end{figure}

To elucidate physics standing for this hysteresis behavior,  magnetizations of the both sublattices are picked out in Fig.3. The results resemble a spin-orientational phase transition in the Heisenberg antiferromagnet and are interpreted as follows. At strong fields both sublattices are polarized along the field. With its decreasing the bigger spins retain their directions while smaller spins change smoothly their alignment into opposite to gain in an  exchange energy.  During further demagnetization  process, when $h<0$,  the bigger spins reorient to be again arranged along the field. This causes a sharp spin reorientation in another sublattice due to the exchange coupling which is  stronger than the corresponding Zeeman interaction. The rapid change is accompanied by the side effect which stands out more noticeably in the MFA calculations, namely, an appearance of "whiskers" in the entire hysteresis curve at low temperatures.  The process is completed by a  gradual saturation of small-spin sublattice magnetization along the applied field. The back sweep goes on similarly.

\begin{figure}
	\begin{center}
		\includegraphics[width=135mm]{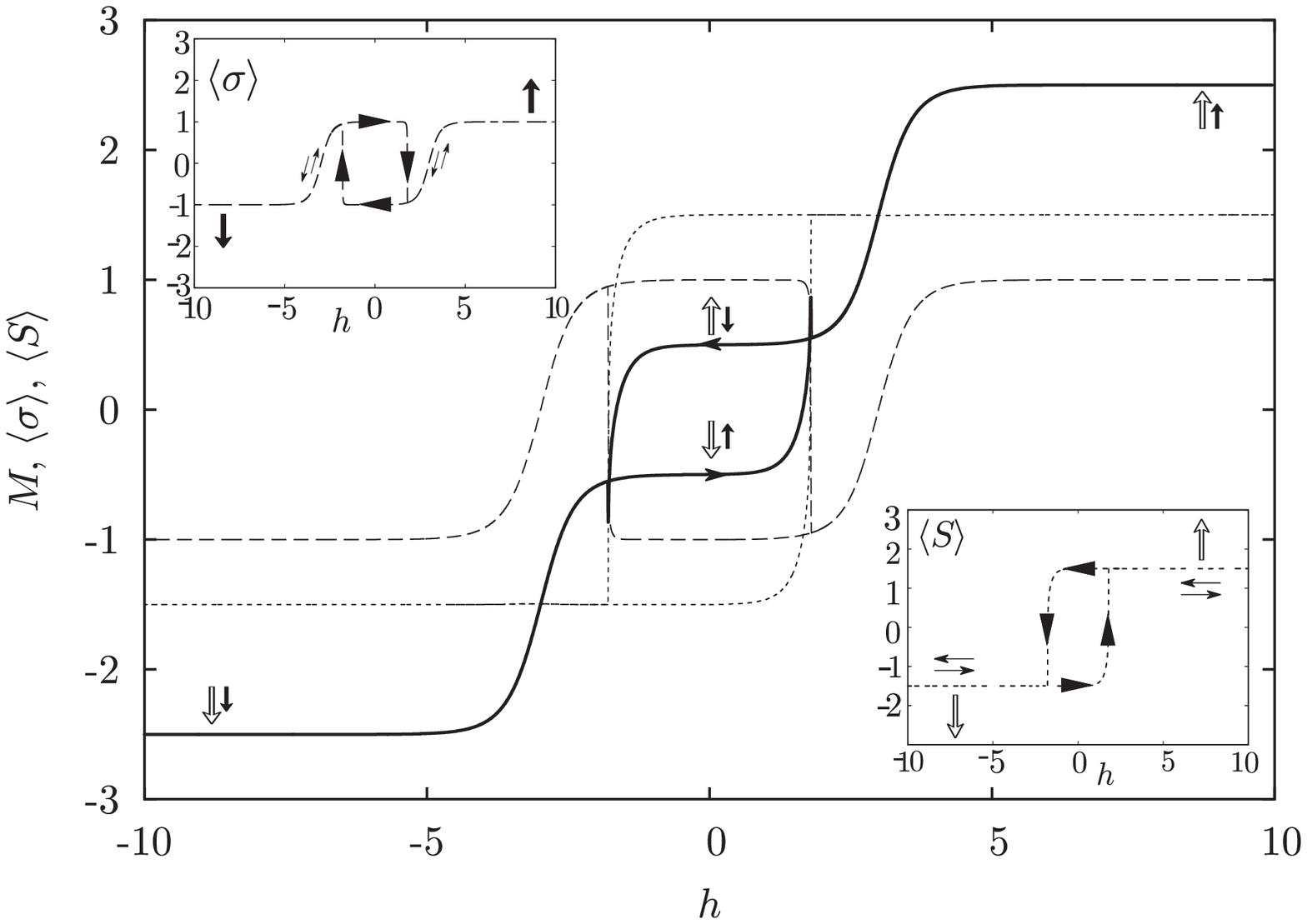}
	\end{center}	
		\caption{Hysteresis loops for the sublattice  $\langle \sigma \rangle$ (big dotted),  $\langle S \rangle$ (small dotted) and the entire  $M$ (solid) magnetizations ($\theta=0.4$).}
	\label{fig:sublattices}
\end{figure}

We also found out how the magnetization process depends on the realtionship between the frequency $\Omega$ (a number of spin transitions per unit time) and the magnetic field frequency $\omega$. We plotted the corresponding curves in Fig.4 where two cases are presented, namely, the quasistatic regime $\omega/\Omega = 10^{-4}$ and the regime when the discrepancy between both the frequencies are not so drastic $\omega/\Omega = 10^{-1}$. As well we show in the Figure the pure static magnetization process which  almost coincides with  that of the quasistatic regime.
 One see that in the case $\omega/\Omega = 10^{-1}$ the hysteresis loop transforms into the narrow S-like form similar to those found experimentally  in the  compound  CoPhOMe at $2.0$ K and $3.5$ K. Physically, the increasing the ratio $\omega/\Omega$ at a fixed magnetic field frequency means that a spin dynamics governed by $\Omega$ slows down.
\begin{figure}
	\begin{center}
		\includegraphics[width=135mm]{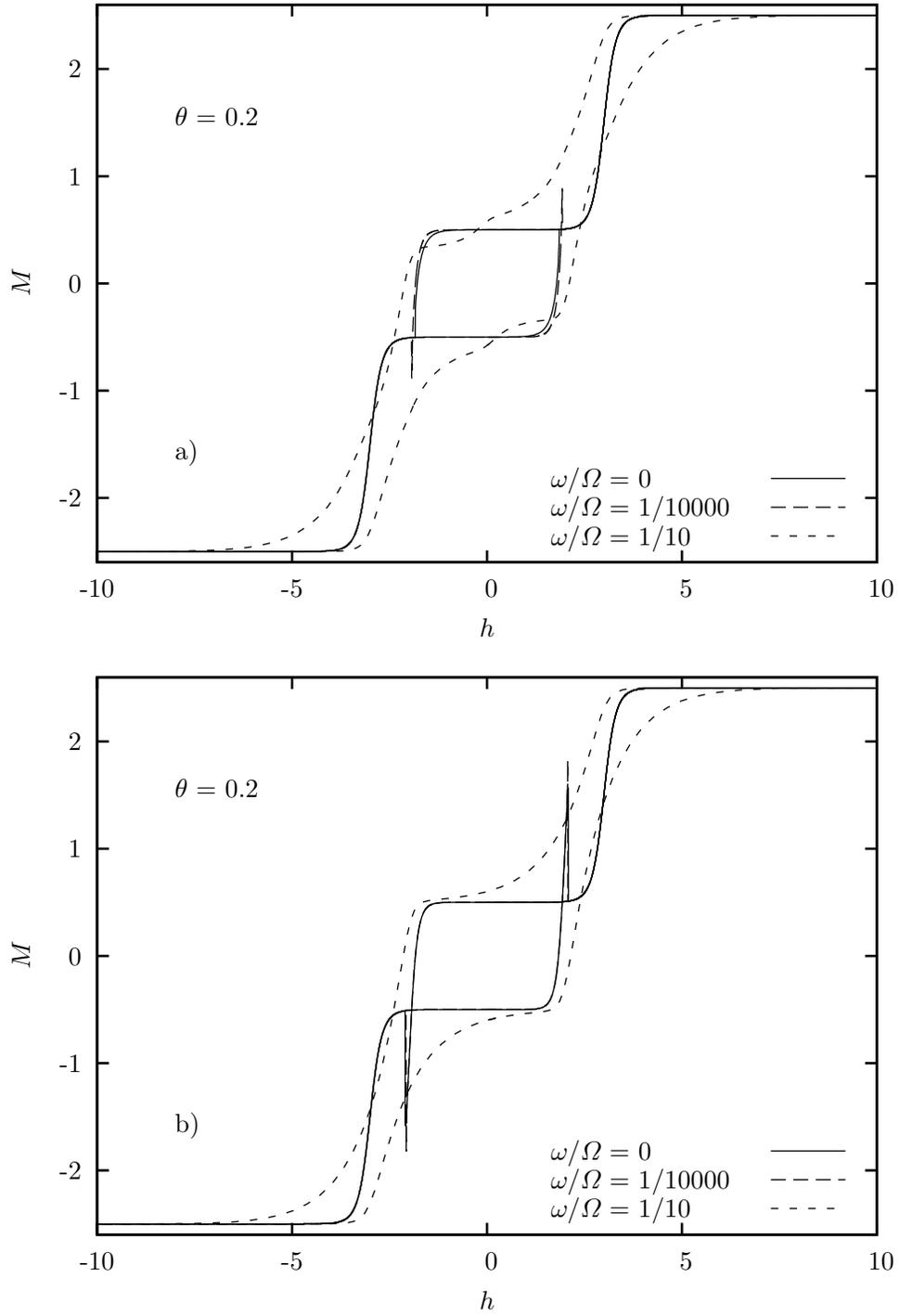}
	\end{center}	
		\caption{Evolution of magnetization curves for different frequency regimes ($\theta=0.2$): (a) MFA; (b) generalized MFA}
	\label{fig:finitesize}
\end{figure}

\section{Confirmation by Monte Carlo simulations}

In the previous section we have shown that in the quasistatic regime $\omega \ll \Omega$  the both analytical models demonstrate  an appearance of hysteresis loops of a peculiar form in the middle of  magnetization curves.  However, we note that spin fluctuations taken into account in the framework the generalized MFA are strongly restricted since they hold a translational symmetry of the chain, i.e. they are the same within an  each elemenatry cell. In order to check whether the hysteresis phenomena are stable against fluctuations in common case, we study the static magnetization process by a Monte Carlo method. 

We apply standard importance sampling methods\cite{Cardona,Newman} to simulate the Hamiltonian given by Eq.(\ref{Hamiltonian}).  Periodic boundary conditions on $N$ = 64, 256 chains were imposed and configurations were generated by sequentially traversing the sublattices and making single-spin flip attempts. The flips are accepted or rejected according to a heat-bath algorithm. Our data were generated with 10$^4$ Monte Carlo steps per spin in the chain after 10$^3$ warming steps  per spin. We checked that the size effects do not substantially effect the results (Fig. 5). 

\begin{figure}[h]
	\begin{center}
		\includegraphics[width=115mm]{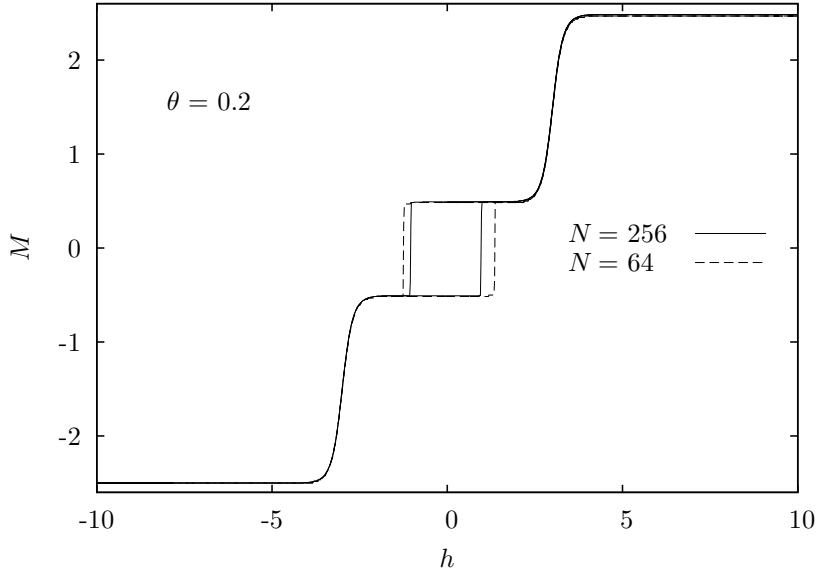}
\end{center}		
		\caption{MC magnetization curves for different chain lengths.}
	\label{fig:finitesize}
\end{figure}

Qualitatively, the same features of the hysteresis loops as obtained in MFA and the generalized MFA are also found in MC. Namely, the hysteresis loop of an almost ideal rectangle form appears in the middle of the magnetization curve at low temperatues. The temperature evolution of the hysteresis curves  reproduces qualiatatively that of predicted by the analytical treatments, however, the MC hysteresis loop narrows more rapidly with an increasing of temperature (Fig. 6).  

\begin{figure}[h]
	\begin{center}
		\includegraphics[width=135mm]{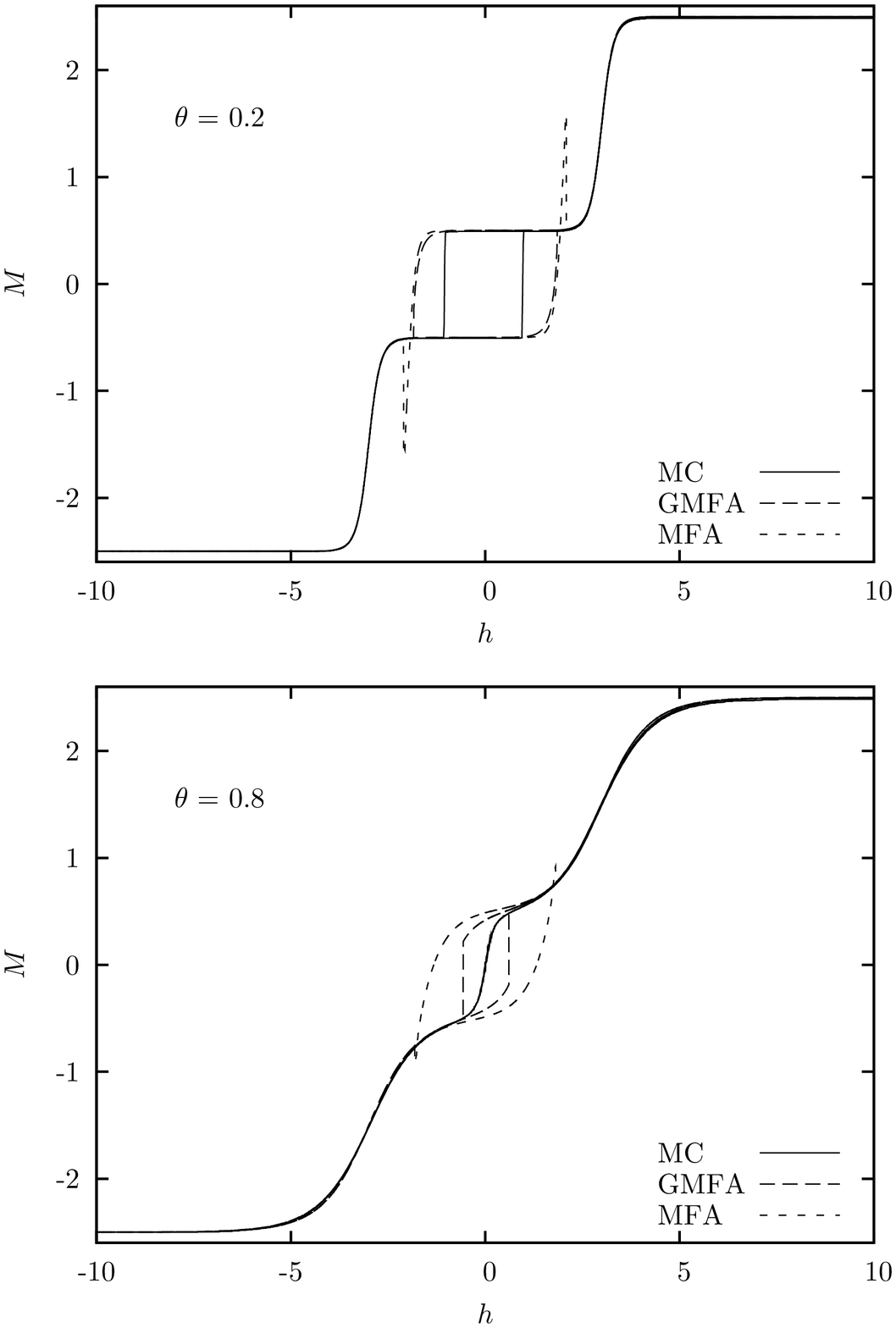}
\end{center}		
		\caption{Temperature evolution of MC hysteresis loops in a comparison with the mean field (MF) and the generalized  mean field (GMF) calculations ($N=256$).}
	\label{fig:finitesize}
\end{figure}

\section{Conclusions}
We analyse a magnetic hysteresis of Co-based quasi-1D ferrimagnetic magnets within the model of  the mixed spin Ising chain. By using a Glauber dynamics approach we build hysteresis loops that come up when a  sinusoidal magnetic field is applied. We found that the unusual shape of the calculated hysteresis cycles coincide with those found experimentally in the CoPhOMe ferrimagnet at low temperatures. However, another Co-based molecular chains\cite{Ishii,Baranov} demonstrate a hysteresis behavior that is scarcely agreed with the analytical treatment, i.e. these materials behave similar to very hard magnets with a high coercive field and broad hysteresis loops. 

It is likely that a reason behind the hardness of the  materials is interchain interactions that result in 3D ordering at  low temperatures, and, as consequence, change a character of elementary excitations. Namely, random spin flips throughout the Ising chain, that is a feature of the Glauber dynamics, are substituted for spin flips in the vicinity of domain walls  separating regions of opposite magnetizations that causes their displacements. Then the demagnetization process is related with the thermally activated DW motion. Similar ideas have been recently argued in Ref.\cite{Sessoli}, where the Mn(III)-based systems\cite{Bernot,Coulon} were suggested to find out a role of easy axis anisotropy in hysteresis phenomena.

\end{document}